\documentclass{article}

\usepackage{arxiv}

\usepackage{amsmath}
\usepackage[utf8]{inputenc} 
\usepackage[T1]{fontenc}    
\usepackage{hyperref}       
\usepackage{url}            
\usepackage{booktabs}       
\usepackage{amsfonts}       
\usepackage{nicefrac}       
\usepackage{microtype}      
\usepackage{cleveref}       
\usepackage{lipsum}         
\usepackage{graphicx}
\usepackage{natbib}
\usepackage{doi}

\title{Reconstructing the Tropical Pacific Upper Ocean using Online Data Assimilation with a Deep Learning model}

\date{}

\newif\ifuniqueAffiliation
\uniqueAffiliationtrue

\ifuniqueAffiliation 
\author{ \href{https://orcid.org/0000-0001-5706-592X}{\includegraphics[scale=0.06]{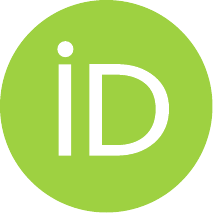}\hspace{1mm}Zilu Meng}\thanks{Corresponding author: Zilu Meng, zilumeng@uw.edu} \\
	Department of Atmospheric Sciences\\
	University of Washington\\
	Seattle, WA 98125 \\
	\texttt{zilumeng@uw.edu} \\
	\And
	\href{https://orcid.org/0000-0001-8486-9739}{\includegraphics[scale=0.06]{orcid.pdf}\hspace{1mm} Gregory J. Hakim} \\
	Department of Atmospheric Sciences \\
	University of Washington\\
	Seattle, WA 98125 \\
	\texttt{ghakim@uw.edu} \\
}
\else
\usepackage{authblk}

\fi

\renewcommand{\headeright}{}
\renewcommand{\undertitle}{}
\renewcommand{\shorttitle}{manuscript submitted to \textit{Journal of Advances in Modeling Earth Systems (JAMES)}}

\usepackage{xcolor}
\hypersetup{
colorlinks,
linkcolor={red!50!black},
citecolor={blue!50!black},
urlcolor={blue!80!black}
}

\begin{document}
\maketitle

\begin{abstract}
A deep learning (DL) model, based on a transformer architecture, is  trained on a climate-model dataset and compared with a standard linear inverse model (LIM) in the tropical Pacific. We show that the DL model produces more accurate forecasts compared to the LIM when tested on a reanalysis dataset. We then assess the ability of an ensemble Kalman filter to reconstruct the monthly-averaged upper ocean from a noisy set of 24  sea-surface temperature observations designed to mimic existing coral proxy measurements, and compare results for the DL model and LIM. Due to signal damping in the DL model, we implement a novel inflation technique by adding noise from hindcast experiments. Results show that assimilating observations with the DL model yields better reconstructions than the LIM for observation averaging times ranging from one month to one year. The improved reconstruction is due to the enhanced predictive capabilities of the DL model, which map the memory of past observations to future assimilation times.
\end{abstract}

\section*{Plain Language Summary}

We use a deep learning (DL) model to better predict climate patterns in the tropical Pacific upper ocean, and to reconstruct past conditions from a sparse network of noisy observations. The DL model forecasts are more accurate than a reference Linear Inverse Model (LIM), which has approximately comparable computational demand. After we adjust DL model forecasts to better approximate errors, we show that this model can more accurately reconstruct climate fields than the LIM. This success highlights the significant potential of deep learning to improve our understanding and prediction of climate change through reconstructing climate variables from sparse information such as from coral proxies.


\section{Introduction}

Owing to the limited time span of satellite observations, our understanding of climate variability on interdecadal and longer timescales derives mainly from climate model simulations and reconstructions from paleoclimate proxies. For example, the El Niño Southern Oscillation (ENSO), one of the most significant drivers of interannual variability in the Earth's system, has a profound impact on the global climate and robust teleconnections
\citep{caneExperimentalForecastsNino1986,mcphadenENSOIntegratingConcept2006b,timmermannNinoSouthernOscillation2018}. However, due to the short period of satellite observations, our understanding of ENSO variability, such as in its intensity, spatial distribution of temperature anomalies, and its interactions with other climate phenomena, is uncertain and poorly sampled \citep{darrigoVariabilityENSOSix2005,ashokNinoModokiIts2007,kugTwoTypesNino2009}. To address this issue, a longer time span of data is needed to better understand these variations in ENSO \citep{timmermannNinoSouthernOscillation2018}. Here we assess the potential of a deep-learning model of the tropical Pacific ocean for assimilating a sparse network of noisy sea-surface temperature (SST) observations, with the goal of reconstructing past climate states from coral proxies of SST. This work provides proof of concept for this goal, and a benchmark comparison to similar approaches using linear inverse models (LIMs).

Reconstructions using paleoclimate proxy data \citep[e.g.,][]{mannGlobalscaleTemperaturePatterns1998,demaesschalckMahalanobisDistance2000a}, such as tree rings, ice core, and coral archives provide insights into past climate conditions. A major challenge with these data sources is their uneven spatial and temporal distribution, which complicates interpretation of signals in climate studies based on multiple proxies. To overcome this issue, the method of climate field reconstruction (CFR) has been used to combine information from different proxies \citep[e.g.,][]{mannGlobalscaleTemperaturePatterns1998}. The primary objective of this approach is to use statistical methods to reconstruct regularly gridded climate fields from sparse and unevenly distributed paleoclimate data. This facilitates studies of global or regional climate variability and may significantly aid in understanding and predicting future climate change.

Recent studies have employed an objective framework for CFR based on data assimilation (DA) \cite[e.g.,][]{goosseReconstructingSurfaceTemperature2010,hakimLastMillenniumClimate2016, vallerModERAGlobalMonthly2024}. One notable difference in the DA approach to climate reconstruction as compared to weather prediction concerns the use of a model to generate the prior (``first guess") before assimilation. A key challenge is that the high operational costs of climate models render ensemble forecasting impractical, which has led to the use of ``offline" DA approaches that randomly sample existing climate model simulations. This approach of offline assimilation has considerable advantages: it requires dramatically lower computational costs; random sampling from the outputs of climate model runs allows for a better estimation of the uncertainty in the assimilation results; and it can yield better outcomes, especially when the predictive capacity of climate models is limited.

Progress toward computationally feasible online paleoclimate DA has been demonstrated using a Linear Inverse Model (LIM) \citep{penlandPredictionNinoSea1993} as an emulator for climate models in the forecast step \citep{perkinsReconstructingPaleoclimateFields2017,perkinsLinearInverseModeling2020}. This approach  transfers climate information from one time step to the next, provides superior priors and more effective use of sparse proxy data. This improvement in assimilation is primarily attributed to the coupled dynamics of the ocean--atmosphere system \citep{perkinsReconstructingPaleoclimateFields2017}, since the largest sources of proxy data, such as tree rings and ice cores \citep{pages2kconsortiumGlobalMultiproxyDatabase2017}, are primarily located on continents and largely reflect atmospheric variations. Using DA with a skillful coupled atmosphere--ocean model allows for this atmospheric information  to inform oceanic state estimates. Given that the predictability of the ocean is substantially higher than that of the atmosphere, using the LIM effectively transmits information through the ocean's memory onto the atmosphere, which benefits proxy assimilation at later times.

Recently, the emergence of deep learning provides a new approach \citep[e.g.,][]{lecunDeepLearning2015a, reichsteinDeepLearningProcess2019} to computationally efficient online data assimilation. Through complex network architectures, deep learning can fit nonlinear relationships in data, thereby enabling more accurate predictions of future states. Moreover, many deep learning models have demonstrated predictive capabilities that exceeding those of LIMs, and even traditional numerical climate models \citep{hamDeepLearningMultiyear2019,zhouSelfattentionBasedNeural2023}. For instance, simple Convolutional Neural Networks (CNNs) have shown remarkable success in predicting the time series of ENSO events with 17 months lead time \citep{hamDeepLearningMultiyear2019}, surpassing the best climate models. However, neural networks like these are designed to predict a single variable, not a field of variables. Therefore, to predict spatial fields for data assimilation, we need to select network architectures that are field-to-field. Recently, the work by \cite{zhouSelfattentionBasedNeural2023} introduced a neural network based on a self-attention structure \citep{vaswaniAttentionAllYou2017}, which they call Geotransformer. The Geotransformer effectively forecasts the monthly ocean temperature and surface wind stress fields of the tropical Pacific to produce ENSO forecasts having skill comparable to those of the aforementioned CNNs. Here we test the use of a version of this model in forecasting and data assimilation experiments to provide proof-of-concept for use in paleoclimate DA. Although this work is motivated by paleoclimate data assimilation, we note that the methods outlined here are also broadly applicable to assimilating instrumental observations.

The remainder of the paper is organized as follows. Section \ref{data_method} details the data and methodologies employed in the construction of the LIM and DL models, as well as the DA method. Section \ref{obsnetwork} delineates the sparse observational network utilized for the DA experiments. Comparative analyses of forecasting performance between the LIM and DL models are presented in Section \ref{prediction_skill}, with the DA experiments elaborated in Section \ref{dl_res}. Finally, Section \ref{condis} offers a discussion of the findings and draws conclusions.

\section{Data, Models, and Data Assimilation Methods} \label{data_method}

Here we provide a detailed description of the data and methods used in our study, including the theory and training procedures for the models, data assimilation techniques, observations, and methods to address the loss of ensemble variance in DL model forecasts.

\subsection{Data}

The focus of our research is the Tropical Pacific (seeing in Figure \ref{fig:proxy}), with an emphasis on the dynamics of ENSO, which is the dominant source of annual to interannual variability in this region \citep{caneExperimentalForecastsNino1986, timmermannNinoSouthernOscillation2018}. Consequently, we adopt variables that are integral to understanding ENSO dynamics, namely sea surface temperature (SST), surface wind stress, and the temperature of the upper ocean on 7 constant layers ranging from 5m to 150m depth (5m, 20m, 40m, 60m, 90m, 120m and 150m). The geographical scope of the data extends from 90°E to 30 °W and 20°N to 20°S. The zonal grid resolution is 2° and the meridional grid resolution is 0.5° (1°) between (poleward of) 5°S and 5°N.

We train our models on data from the Coupled Model Inter-comparison Project Phase 6 (CMIP6) historical experiments \citep{oneillScenarioModelIntercomparison2016a}, and initialize and verify results on data from the Simple Ocean Data Assimilation Products (SODA) \citep{cartonReanalysisOceanClimate2008} and the Global Ocean Data Assimilation System (GODAS) reanalysis. We note that the reanalysis data period spanned by combining GODAS and SODA covers a span of only about 130 years, which is somewhat limited for training a neural network with a large number of parameters. Consequently, we rely on the CMIP6 model data for training our network, consisting of 23 CMIP6 models (Figure \ref{fig:nino34_scale}). We then use SODA data for validation and to fine-tune the network architecture and hyper-parameters (described in the section \ref{sec:training}). The GODAS dataset is used as the final test set for assessing model performance. In this context, the training, validation, and testing datasets are derived from different sources, which helps to prevent overfitting and independently assess results. Summary information about these datasets is provided in Table \ref{table1}.

\begin{table}
  \centering
  \caption{Dataset and Source}
  \begin{tabular}{lccc}
    \toprule
    \textbf{Type} & \textbf{Source} & \textbf{Period} \\
    \midrule
    Train set &\begin{tabular}{c}Coupled Model Inter-comparison \\ Project phase 6 (CMIP6)\end{tabular}  & Jan. 1850 to Dec.2014 \\
    Validation set &  \begin{tabular}{c}Simple Ocean Data Assimilation \\  (SODA) products\end{tabular} & Jan. 1871 to Dec. 1979 \\
    Test set & \begin{tabular}{c}Global Ocean Data Assimilation \\ System (GODAS) reanalysis \end{tabular} & Jan. 1980 to Dec. 2021 \\
    \bottomrule
  \end{tabular}
  \label{table1}
\end{table}

\subsection{Models}

\subsubsection{Linear Inverse Model (LIM)}

Linear Inverse Modeling (LIM) is an efficient, widely applied, and powerful model for sea surface temperature prediction and assimilation, especially for ENSO \citep[e.g.,][]{penlandPredictionNinoSea1993,penland1995optimal,newmanEmpiricalBenchmarkDecadal2013}. The LIM is an empirically determined estimate of a dynamical system linearized about its mean state:

\begin{equation}
    \frac{d \boldsymbol{x}}{dt} = \mathbf{L} \boldsymbol{x} + \boldsymbol{\xi},
    \label{eqn:LIM}
\end{equation}
in which $\boldsymbol{x}$ is the state vector, typically cast in terms of a truncated set of Empirical Orthogonal Functions (EOFs), $\mathbf{L}$ is the linear system operator, and $\boldsymbol{\xi}$ represents noise that is white in time but correlated in the state variables. $\mathbf{L}\boldsymbol{x}$ represent the deterministic tendency of the system, and $\xi$ a stochastic forcing that represents the net effect of nonlinearity and unresolved processes. Integrating (\ref{eqn:LIM}) over $t=0:\tau$, for any $\tau$, and taking the expected value, gives
\begin{equation}
    \boldsymbol{x}(\tau) = \mathbf{G}_{\tau} \boldsymbol{x}(0),
\end{equation}
\noindent where $\mathbf{G}_{\tau} = exp(\mathbf{L}\tau)$. Given sample training data, $\mathbf{G}_{\tau}$ may be determined for a single $\tau$ from regular least-squares regression:

\begin{equation}
    \mathbf{G}_{\tau} = \mathbf{C}(\tau) \mathbf{C}(0)^{-1}.
    \label{eq:lim_train}
\end{equation}
\noindent Here $\mathbf{C}(\tau)$ is the lag covariance matrix of the state vector at lag time $\tau$, 
$$
\mathbf{C}({\tau}) = <\boldsymbol{x}^T(\tau) \boldsymbol{x}(0)>
$$ 
\noindent and ``$<>$" represents a sample average. 
Matrix $\mathbf{L}$ is then determined from $\mathbf{G}_{\tau}$. The stochastic noise $\xi$ is assumed to be a white noise process with a covariance matrix $\mathbf{Q}$, meaning $ <\boldsymbol{\xi}^T \boldsymbol{\xi}> = \mathbf{Q} $. Assuming stationary statistics, matrix $\mathbf{Q}$ is constant and defined by 
\begin{equation}
    \frac{d \mathbf{C}(0)}{dt} = \mathbf{L} \mathbf{C} (0) +  \mathbf{C} (0) \mathbf{L}^{T} + \mathbf{Q} = 0.
    \label{eq:dcdt0}
\end{equation}

Forecasts are computed using 
\begin{align}
    \boldsymbol{x}_{t+\delta t} = (\mathbf{L} \delta t + \mathbf{I})\boldsymbol{x}_{t} + \boldsymbol{\hat{Q}} \sqrt{\mathbf{\Lambda} \delta t} \boldsymbol{\alpha} \\
    \boldsymbol{x}_{t + \delta t / 2} = \frac{1}{2}  (\boldsymbol{x}_{t+\delta t} +  \boldsymbol{x}_{t}),
\end{align}
in which $\boldsymbol{\alpha}$ is a vector of independent standard normal random variables, $\mathbf{\Lambda}$ and $\mathbf{\hat{Q}}$ are the eigenvalues and eigenvectors of $\mathbf{Q}$. In this study, $\delta t$ is set to 6 hours. In this context, the LIM can be viewed as a linear system driven by spatially correlated, temporally white, noise, representing nonlinear and unresolved fast processes. More details of the LIM can be found in \cite{penlandPredictionNinoSea1993} and \cite{newmanEmpiricalModelTropical2011}.



\subsubsection{Deep Learning Model (DL)}

\citet{zhouSelfattentionBasedNeural2023} introduce a novel self-attention-based neural network specifically designed for predicting the tropical Pacific upper ocean. The model is structured to take a 12-month state vector as input, with fields including SST, surface wind stress, and upper ocean temperature on 7 vertical levels. The output of this model is a state vector for the subsequent 12 months, which can be regressively extended indefinitely into the future:
\begin{equation}
\boldsymbol{X}_{t+1:t+12}^{\text{out}} = \boldsymbol{DL}(\boldsymbol{X}_{t-12:t}^{\text{in}}),
\end{equation}
\noindent in which $\boldsymbol{X}_{t-12:t}^{\text{in}}$ is the input state vector from time $t-12$ to $t$, $\boldsymbol{X}_{t+1:t+12}^{\text{out}}$ denotes the forecasted state vector covering the period from time $t+1$ to $t+12$, and $\boldsymbol{DL}$ is the deep learning model operator. 

This model initially employs an embedding layer to encode the data into smaller blocks based on latitude and longitude, into a vector of length 256. Subsequently, it utilizes temporal self-attention and spatial self-attention modules to extract features. Finally, the model outputs the predicted results through a fully connected layer. More details of the model can be found in \cite{zhouSelfattentionBasedNeural2023}.

Compared with the LIM, there are two main advantages of DL model. First, the DL model can capture the  deterministic non-linear relationships between the current and future states. Second, the DL model employs nonlinear dimensional reduction, which retains more predictive capability. As we will show, the DL model has better prediction skill than the LIM (see Section \ref{prediction_skill} for details).

\subsubsection{Model Training and Configuration}\label{sec:training}

\textbf{LIM}. The first step in LIM training employs multivariate Empirical Orthogonal Functions (EOFs) \citep[e.g.,][]{lorenz1956empirical, hannachi2007empirical}, meaning applying EOFs to the covariance matrix for each variables and truncating to reduce the dimensionality of the three variables mentioned above (SST, wind stress and ocean temperature). All variables are standardized by dividing them by their respective standard deviations before performing the EOF calculation. We then use dimensionally reduced Principal Component (PC) time sequences of these variables to construct the LIM from equation (\ref{eq:lim_train}). To ensure that LIM achieves its best forecasting ability, we performed dimensional reduction on the data from each of the 23 CMIP6 models and conducted an exhaustive search to identify the optimal number of PCs for training a LIM from each model. The search objective is to identify the highest 12-month mean Nino3.4 Index forecast correlation skill for each model. We find that the first 12 PC sequences from the GFDL-CM4 model yield the best forecasting results of the 23 LIMs evaluated. Therefore, we use the first 12 PC sequences from the GFDL-CM4 model \citep{adcroftGFDLGlobalOcean2019} to build our LIM. It is noteworthy that previous studies have constructed LIMs using only SST or a combination of SST and Sea Surface Height (SSH), rather than also predicting the mixed layer as we do here. Thus, to provide a basis for comparison with previous work, we also construct a LIM using only SST, which we refer to as LIMOsst. A comprehensive search again reveals that the LIM built with the first 12 PC sequences of GFDL-CM4 SST yields the best forecasting performance.

\textbf{DL Model}. As our objective is to predict the state of the entire field for the next time step, we have modified the loss function of the original model described in \citep{zhouSelfattentionBasedNeural2023}. Instead of incorporating the Root Mean Squared Error (RMSE) of the Nino3.4 index into the loss function as in the original study \citep{zhouSelfattentionBasedNeural2023}, we have adopted the RMSE of the entire state vector as our loss function:
\begin{equation}
        \text { Loss }=\frac{1}{T_{\text {out}}} \sum_{t=1}^{T_{\text {out }}} \sqrt{\frac{1}{N_{\text {lat}} \times N_{\text {lon }} \times C} \sum_{i=1}^{N_{\text {lon}}} \sum_{j=1}^{N_{\text {lat}}} \sum_{k=1}^C\left(x_{t, k , j , i}^{\text {out}}-x_{t, k, j, i}^{\text{true}}\right)^2},
\end{equation}
\noindent where $x^{\text{out}}$ represents the forecast states, while $T_{\text{out}}$, $N_{\text{lat}}$, and $N_{\text{lon}}$ denote the number of output time steps, the number of latitude grids, and the number of longitude grids, respectively.

For training we use the PyTorch framework \citep{paszke2019pytorch} and source code from \citet{zhouSelfattentionBasedNeural2023}. To achieve optimal tuning of the network parameters, we use the Adam optimization algorithm \citep{kingma2014adam}, implementing a decaying learning rate strategy that starts at 0.0005 and decreases with ongoing training. Moreover, we incorporate an early stopping mechanism \citep{prechelt2002early} that halts training if there is no reduction in the validation RMSE over four consecutive epochs. All training strategies are detailed with the code accompanying this study (see section \ref{sec:codedata}).

\subsection{Data Assimilation Methods}

This section describes the assimilation methods used in this study, including augmenting the error of the DL model ensemble forecasts.

\subsubsection{Ensemble, Online and Offline Assimilation\label{EnSRF}}

We perform data assimilation using an Ensemble Kalman Filter (EnKF) \citep{evensenDataAssimilationEnsemble2009}, which has been shown to perform well in paleo-data assimilation tasks \citep{zhuPseudoproxyEmulationPAGES2023,hakimLastMillenniumClimate2016,perkinsReconstructingPaleoclimateFields2017,tardifLastMillenniumReanalysis2019}. The first part of the Kalman filter is the update step:  

\begin{equation}
\boldsymbol{x}_{a}=\boldsymbol{x}_{p}+\mathbf{K}\left[\boldsymbol{y}-\mathcal{H}\left(\boldsymbol{x}_{\boldsymbol{p}}\right)\right],
\label{eqn:update}
\end{equation}
\noindent in which $\boldsymbol{x}_{a}$ is the analysis state vector, $\boldsymbol{x}_{p}$ is the prior state vector, $\boldsymbol{y}$ is the observation vector, $\mathcal{H}$ is the observation operator, and $\mathbf{K}$ is the Kalman gain matrix defined by:

\begin{equation}
    \mathbf{K}=\mathbf{B H}^{\mathbf{T}}\left[\mathbf{H B H ^ { \mathbf { T } }}+\mathbf{R}\right]^{-1},
    \label{eqn:gain}
\end{equation}
\noindent Here, $\mathbf{B}$ is the prior covariance matrix, $\mathbf{H}$ is a linearized version of $\mathcal{H}$, and $\mathbf{R}$ is the observation error covariance matrix. In the Last Millennium Reanalysis \citep[LMR]{hakimLastMillenniumClimate2016} framework, $\mathbf{R}$ is a diagonal matrix with diagonal elements equal to the observational error variance. In this study, we use the Ensemble Square Root Filter (EnSRF) method \citep{tippettEnsembleSquareRoot2003} to solve equation (\ref{eqn:update}) and equation (\ref{eqn:gain}), including serial observation processing. Since the covariance spatial length scales on annual to monthly time scales in this region are relatively long, we do not impose covariance localization. The EnSRF method for the $k$th proxy, $y_k$, is described in the following equations. We separate the ensemble into two parts, the ensemble mean and ensemble perturbation. First, we calculate the analysis ensemble mean by:

\begin{equation}
\overline{\boldsymbol{x}_{a}}=\overline{\boldsymbol{x}_{p}}+\frac{\operatorname{cov}\left(\boldsymbol{x}_{p}, y_{e, k}\right)}{\left(\operatorname{var}\left(\boldsymbol{y}_{e, k}\right)+R_k\right)}\left(y_k-\overline{y_{e, k}}\right).
\end{equation}
\noindent Here the overbar ($\overline{x}$) denotes an ensemble mean, and primes ($x^\prime$) denotes an ensemble perturbation. Subscript ``p" denotes the prior state vector (forecast), subscript ``a" the analysis state vector (posterior), $\boldsymbol{y}_{e, k}$ is the $k$th proxy estimate from the ensemble, and $R_k$ is the $k$th proxy error variance. The ``$\operatorname{var}$" and ``$\operatorname{cov}$" are the variance and covariance operators. Second, we calculate the analysis ensemble perturbations from:

\begin{equation}
\boldsymbol{x}_{a}^{\prime}=\boldsymbol{x}_{b}^{\prime}-\left[\mathbf{1}+\sqrt{\frac{R_k}{\operatorname{var}\left(y_{e, k}\right)+R_k}}\right]^{-1} \frac{\operatorname{cov}\left(\boldsymbol{x}_{p}, y_{e, k}\right)}{\left(\operatorname{var}\left(y_{e, k}\right)+R_k\right)}\left(y_{e, k}^{\prime}\right)
\end{equation}

Finally, each ensemble member analysis state vector is calculated by adding the ensemble mean and ensemble perturbation:

\begin{equation}
    \boldsymbol{x}_a = \overline{\boldsymbol{x}_a}+\boldsymbol{x}_a^{\prime}
\end{equation}

One objective of this paper is to demonstrate that low-frequency observations (averaged more than one month) can be effectively assimilated to reconstruct monthly-averaged climate fields. By employing the method mentioned previously, we can update the monthly variables using 3, 6, or 12-month averaged observations through the covariance matrix $\operatorname{cov}\left(\boldsymbol{x}_{p}, y_{e, k}\right)$. 

To assess the impact of the models on the data assimilation results, we perform experiments using both online and offline data assimilation, as subsequently described. All experiments in this paper use 100 ensemble members, a number chosen to balance good results with the constraints of limited computational resources. Results for the DL model improve modestly for larger ensembles (tests for 50--400 members shown in Supplementary Figure 3).

\textbf{Online Assimilation}. After the update step, in the online assimilation method, we perform the forecast step, meaning a forecast initialized with the result of the update step:

\begin{equation}
    \boldsymbol{x}_{p,t+1}= \mathcal{M}(\boldsymbol{x}_{a,t}),
\end{equation}
\noindent where $\mathcal{M}$ represents the model operator. We employ both the LIM and the DL model as the model operators. Additionally, we evaluate the effect of different observation-averaging periods on the analyses, since these vary by climate proxy. Specifically, we consider time averages including  1, 3, 6, and 12 months. Consequently, when assimilating observations, we adapt the model's forward prediction to align with the temporal length of the averaging time of the observations. This allows us to compute the prior values for the proxy data, which are subsequently assimilated using method described above \citep{steigerAssimilationTimeAveragedPseudoproxies2014,huntleyAssimilationTimeaveragedObservations2010}. To initialize the ensemble at the start of each experiment, we randomly select 100 model output data fields from the CMIP6 data that correspond to the current month. After completing assimilation for this randomly drawn ensemble, we make a forecast using this assimilated field as initial condition to the next time for assimilation. This forecast--assimilation cycle continues until the final time with observations.

\textbf{Offline Assimilation}. In the original LMR framework \citep{hakimLastMillenniumClimate2016,tardifLastMillenniumReanalysis2019}, assimilation is executed offline. In the offline case, the prior state vector consistently comprises the same random ensemble drawn from a single CMIP6 model, with no intervening forecast step. To optimize the performance of this offline assimilation, we conduct the assimilation process using data from all 23 CMIP6 models as the prior ensemble and select the most accurate outcome as our final result, as defined by the highest reconstructed Nino3.4 Index correlation. We find that the MPI-ESM2-0 simulation is the optimal source for the offline prior ensemble. This approach ensures that all experiments conducted can validate the relative outperformance of the Deep-Learning--assimilation method for reconstructing tropical Pacific climate fields. 

\subsubsection{DL model Ensemble Inflation}

\begin{figure}
    \centering
    \includegraphics[width=0.9\textwidth]{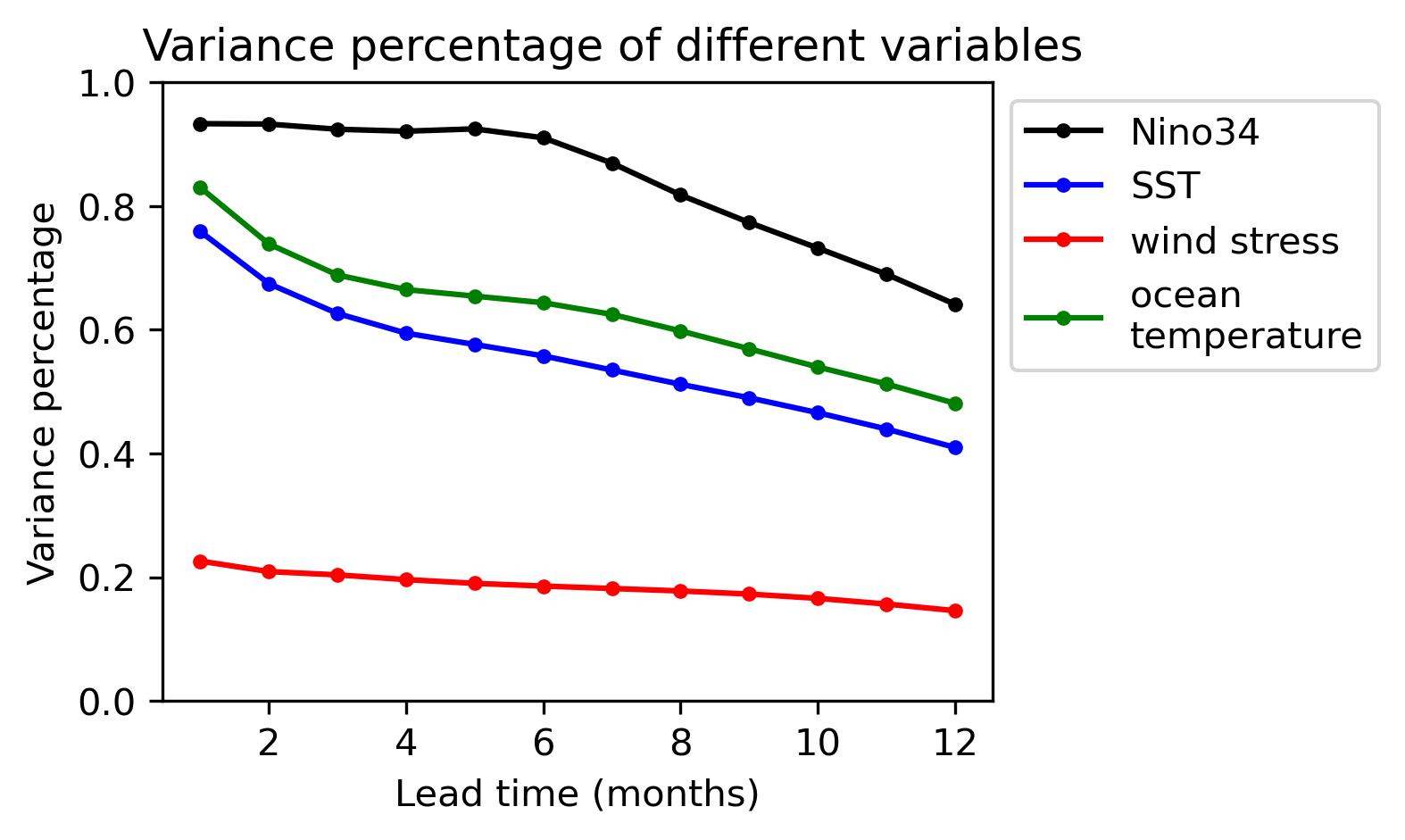}
    \caption{\textbf{Variance proportion of deep learning forecasts compared to GODAS observations across different variables over time.} Illustrated here are the variance ratios for predictions made by the Deep Learning model relative to actual observations from the GODAS dataset, covering variables such as the Nino3.4 Index, Sea Surface Temperature (SST), wind stress, and ocean temperature, as a function of varying lead times in months.}
    \label{precpre}
\end{figure}

In the case of deep learning networks, particularly for tasks involving monthly or annual forecasting, DL models tend to lose error associated with the unpredictable signal. For example, when forecasts are initialized and verified using the GODAS dataset, the ratio of the SST variance between the predictions of the DL model and the target data in the GODAS dataset is less than 1. This ratio decreases with increasing prediction lead time, as illustrated by the blue line in Fig. \ref{precpre}. The variance of the forecast wind stress is also significantly lower than that of GODAS, and the variance ratio of other variables are also similarly smaller. The primary reason for this discrepancy is that the DL model is not trained to capture unpredictable signals, which is especially evident in the atmosphere as seen in supplementary Figure 1. 
This unpredictable noise may be caused by unresolved processes, or by signals that originate outside of the forecast domain, such as the Pacific Meridional Mode (PMM) \citep{vimontSeasonalFootprintingMechanism2003,mengWhyPacificMeridional2024a} from mid-latitudes, the Indian Ocean Dipole (IOD) \citep{sajiDipoleModeTropical1999}, and from other ocean basin and deep-ocean dynamics.

While this is not an issue for forecasting tasks aimed at providing only the predictable future signal, it presents a challenge for data assimilation since it involves estimating errors. Underestimating errors in the forecast can lead to under weighting the information from observations. To address this issue, we  employ a variance inflation technique \citep{evensenDataAssimilationEnsemble2009} by adding random errors  to the DL forecast. Specifically, we assume that the unpredicted error is a random process. We sample from an ensemble of these random vectors obtained by calculating the difference between the DL forecast and the verifying fields from hindcasting experiments. The sample size of GODAS and SODA data is not sufficient to obtain the statistical characteristics of this noise, so we use DL forecast errors from initializing and predicting on the CMIP6 models. Since there is a significant difference between the intensity of ENSO in CMIP6 and observations (SODA) \citep{beobide-arsuagaUncertaintyENSOamplitudeProjections2021}, we use the standard deviation of the Nino3.4 Index (ENSO intensity) as a scaling factor on the errors from CMIP6 models hindcast. Specifically, the random errors added to the DL forecasts are calculated by the following steps.

\begin{figure}
    \centering
    \includegraphics[width=0.9\textwidth]{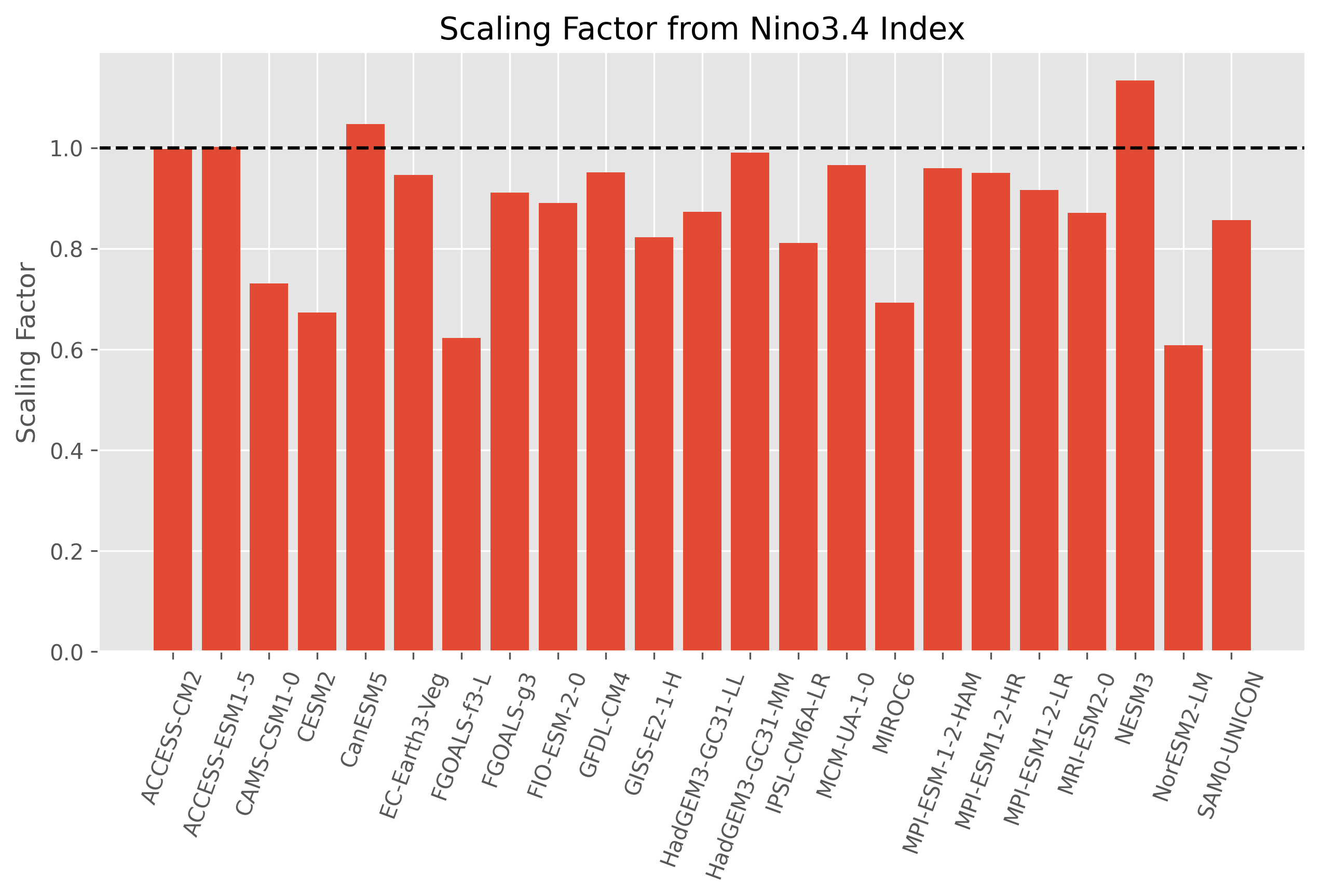}
    \caption{\textbf{Comparison of standard deviation ratios: SODA vs. CMIP6 Models for the Nino3.4 Index.} This graph displays the ratio of the Nino3.4 index standard deviation from the SODA dataset to that of various CMIP6 models, which is utilized as a scaling factor for noise based on the CMIP6 model data.}
    \label{fig:nino34_scale}
\end{figure}

First, we calculate the ratio between the standard deviation of SODA Nino3.4 index and each CMIP6 model's Nino3.4 index as the scaling factor $\alpha_{i}$,  

\begin{equation}
    \alpha_{i} = \frac{\sigma_{\text{SODA}}}{\sigma_{i}}.
    \label{eq:scale}
\end{equation}
\noindent Here, $\sigma_{\text{SODA}}$ is the standard deviation of the SODA Nino3.4 index and $\sigma_{i}$ is the standard deviation of the $i$-th CMIP6 model's Nino3.4 index. The major reason for using Nino3.4 index as scale factor is that the Nino3.4 index standard deviation is representative of the intensity of ENSO. Ratios for most models are less than 1 (Fig. \ref{fig:nino34_scale}), which means the variance of the CMIP6 models ENSO intensity is larger than in SODA. We apply the scaling factor to the forecast errors from CMIP6 models, $\boldsymbol{\eta}_{m,l,i}$, for the $i$th CMIP6 model, $l$th lead time and $m$th ensemble member:

\begin{equation}
    \boldsymbol{\eta}_{m,l,i} = \alpha_{i}(\boldsymbol{x}^{\text{true}}_{m,l,i} - \boldsymbol{x}^{\text{out}}_{m,l,i}).
\end{equation}

The corrected forecast from the DL model is then defined by
\begin{equation}
    \boldsymbol{x}_{t+1} = \boldsymbol{x}^{\text{out}}_{m,l} + \boldsymbol{\eta}_{m,l}.
\end{equation}

We note that, in comparison to adding noise directly, scaling the noise leads to an approximately 5\% improvement in the reconstruction results in terms of correlation.

On the contrary, in the LIM, the variance of the deterministic part of the state vector also decays with time, because the real part of the eigenvalues of $\mathbf{L}$ are less than 0. However, the random noise forcing component of $\boldsymbol{\xi}$ yields an unbiased forecast covariance (as seen in equation \ref{eq:dcdt0}). Therefore, the variance inflation technique is not used for the LIM forecasts.

\subsubsection{Evaluation Criteria}

The major evaluation criteria used in this study for the prediction skill and reconstruction skill are sample time-series correlation and root-mean-squared error (RMSE). The correlation is calculated by the following equation:

\begin{equation}
    \text {corr}=\frac{1}{T} \sum_{t=1}^T \frac{\left(f_t-\bar{f}\right)\left(v_t-\bar{v}\right)}{\sigma_f \sigma_v},
\end{equation}
in which $f_t$ is the prediction result or reconstruction result, $v_t$ is the ground truth at time $t$, $\bar{f}$ and $\bar{v}$ are the mean of prediction result and ground truth, $\sigma_f$ and $\sigma_v$ are the standard deviation of prediction result and ground truth. RMSE is calculated by:
\begin{equation}
    \text{RMSE} = \sqrt{\frac{1}{T} \sum_{t=1}^T (f_t - v_t)^2 }.
\end{equation}

Correlation and RMSE results are averaged over the entire domain to provide summary measures of skill. To quantify the improvement of DL over the LIM and offline DA approaches, we use the improvement ratio (IR):

\begin{equation}
    \text{IR} = \frac{S_{\text{DL}} - S_{\text{traditional}}}{S_{\text{traditional}}},
\end{equation}
in which $S_{\text{DL}}$ and $S_{\text{traditional}}$ are the skill scores of the DL model and the traditional model (the LIM, or the offline method), respectively. The skill score is defined by first computing the domain-averaged correlation or RMSE, and then the IR. For consistency with RMSE, the improvement ratio is multiplied by -1, so that smaller values mean better skill.

\section{Observing Network} \label{obsnetwork}

Here we describe the design of the data assimilation experiments, including the locations of the pseudoproxies, the types of pseudoproxies, the error characteristics for the pseudoproxies, and the experimental setup.

\begin{figure}
    \centering
    \includegraphics[width=0.9\textwidth]{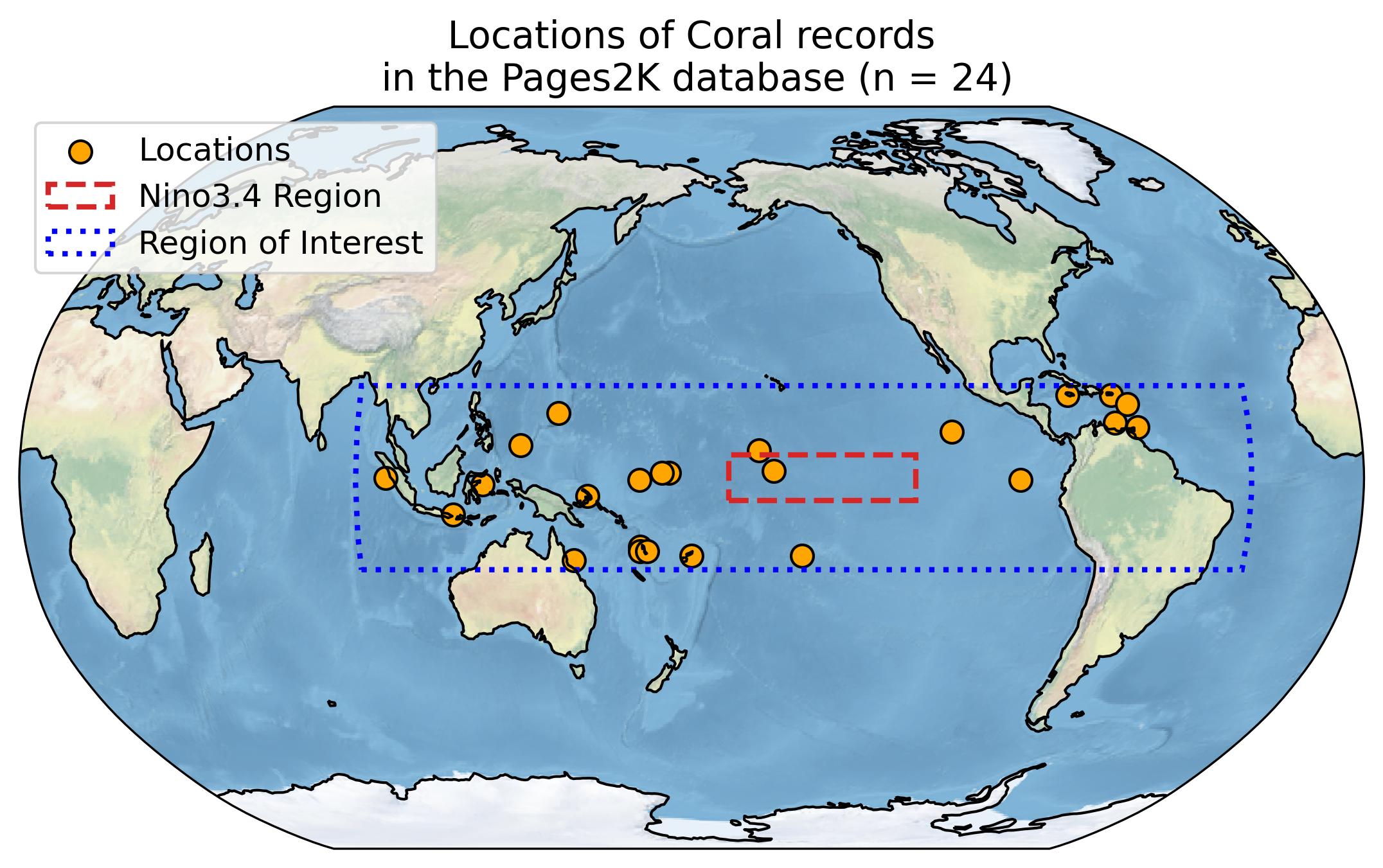}
    \caption{\textbf{Geographical distribution of coral $\delta^{18}O$ proxy records in the tropical Pacific region from the PAGES2K database (Outlined by the dashed blue box).} The dashed blue box delineates the area of focused modeling and interest, the dashed red box indicates the Nino3.4 region, and the brownish-yellow dots represent the locations of coral $\delta^{18}O$ proxy records.}
    \label{fig:proxy}
\end{figure}

To simulate the real assimilation process as closely as possible, we use the locations of stable oxygen isotope composition ($\delta^{18}O$) coral proxy locations from the widely utilized PAGES2K database. We take the average number of coral sites available in the tropical Pacific domain from 1600 to 2000, which amounts to 24, as the locations for the pseudoproxy data (Fig. \ref{fig:proxy}). In the assimilation of paleoclimate data, the proxy data averaging time varies from monthly to annual. Therefore, assimilating these proxies, which represent climatic data over periods longer than a month, to obtain monthly average climate data poses a significant challenge to the forecast model's ability to predict the duration of climate information. To test this capability, we set the duration of the pseudoproxy data averaging time to 1 month, 3 months, 6 months, and 12 months, respectively, and conduct separate assimilation experiments for each duration.

We take observations from the GODAS dataset as ground truth, interpolating to the proxy location and time averaging:

\begin{equation}
    y_{\text{avg}, N} = \frac{1}{N}\sum_{i=k+1}^{k+N} {y_{i}},
\end{equation}
with $N$ taking values of 1, 3, 6, and 12 in this study. To synchronize with ENSO seasonality, we compute 3-month averages corresponding to MAM (March-April-May), JJA (June-July-August), SON (September-October-November), and DJF (December-January-February). The 6-month averages are MAMJJA (March to August) and SONDJF (September to February), while the 12-month average spans MAMJJASONDJF (March to February).

After completing interpolation and averaging, we simulate random errors in the data drawn from a Gaussian distribution with a mean of 0. The Signal-to-Noise Ratio (SNR) \citep{zhuPseudoproxyEmulationPAGES2023}, defined in terms of standard deviation, is set to 1. Additionally, we tested the sensitivity to SNR by conducting additional experiments for SNR = 0.2, 0.5, 2, and 5, and find that the amplitude of the SNR does not significantly impact the results (Supplementary Information). We simulate observation error by

\begin{equation}
    y_{\text{avg}, N} ^{\prime} = y_{\text{avg},N} + \zeta,
\end{equation}
in which $\zeta \sim N(0, \sigma^2)$, and $\sigma$ is the standard deviation of the real data divided by the SNR. The error simulation is performed for each proxy location and each time-averaging duration. 

\section{Forecasting Results \label{prediction_skill}}

\begin{figure}
    \centering
    \includegraphics[width=1\textwidth]{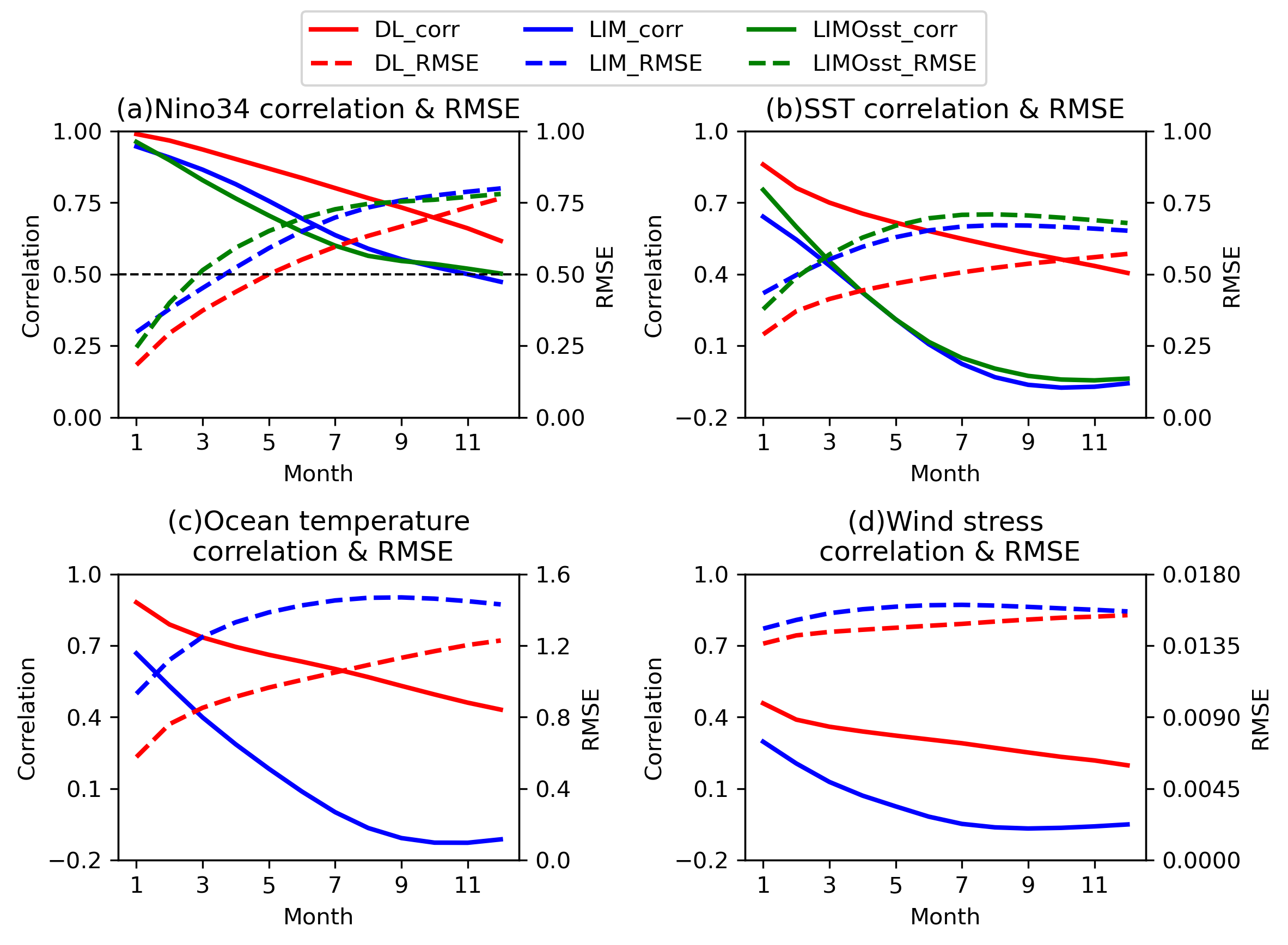}
    \caption{\textbf{Forecast skill of Deep Learning model (red), Linear Inverse Model (blue) \& LIMOsst (green)} in terms of correlation (solid lines) and RMSE (dashed lines) across lead time for the Nino3.4 Index \textbf{(a)}, Sea Surface Temperature (SST) field \textbf{(b)}, ocean temperature field \textbf{(c)}, and the wind stress field \textbf{(d)}. The correlation scale is provided on the left $y$-axis and RMSE on the right $y$-axis, as the function of forecast lead time.}
    \label{fig:predictionskill}
\end{figure}

\begin{figure}
    \centering
    \includegraphics[width=\textwidth]{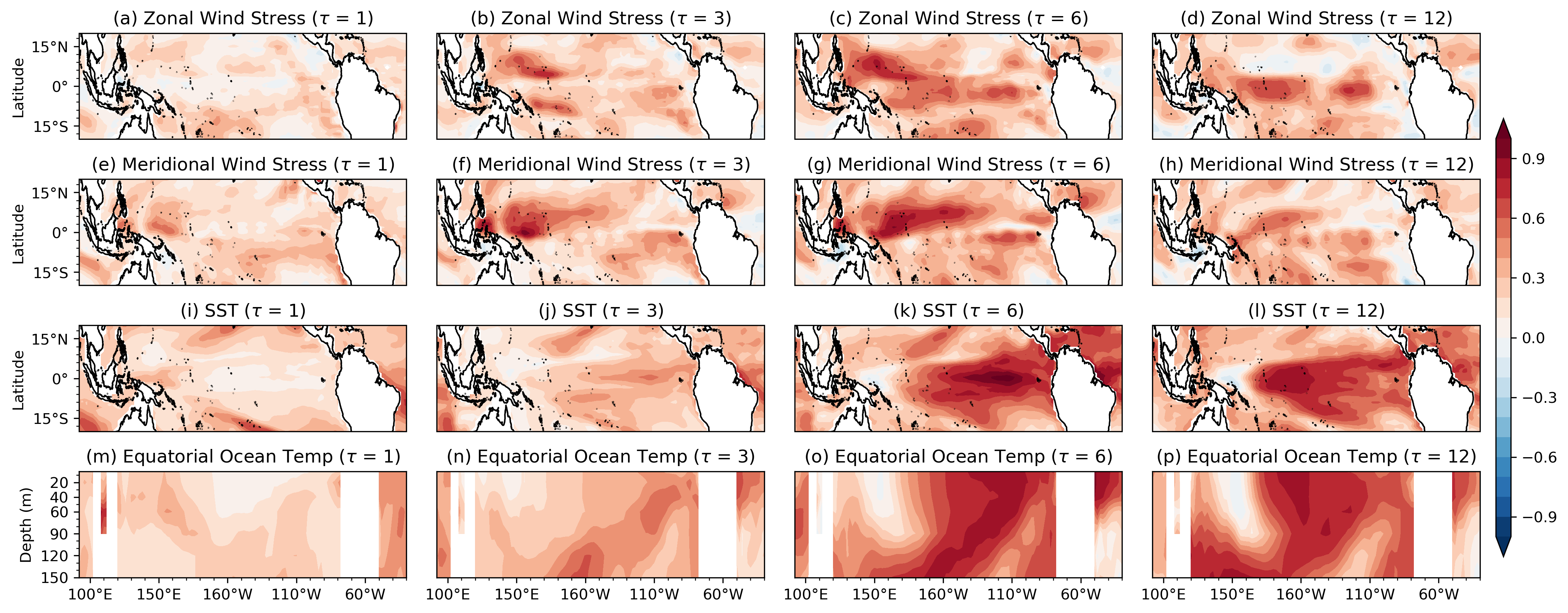}
    \caption{\textbf{Difference in forecast skill as measured by correlation between Deep Learning and Linear Inverse Models as a function of lead time ($\tau$).} The first row (a–d) shows zonal-wind stress, the second row (e–h) meridional wind stress, the third row (i–l) Sea Surface Temperature (SST), and fourth row (m–p) equatorial (5°N–5°S) ocean temperature.}
    \label{fig:predcorrdiff}
\end{figure}

\begin{figure}
    \centering
    \includegraphics[width=\textwidth]{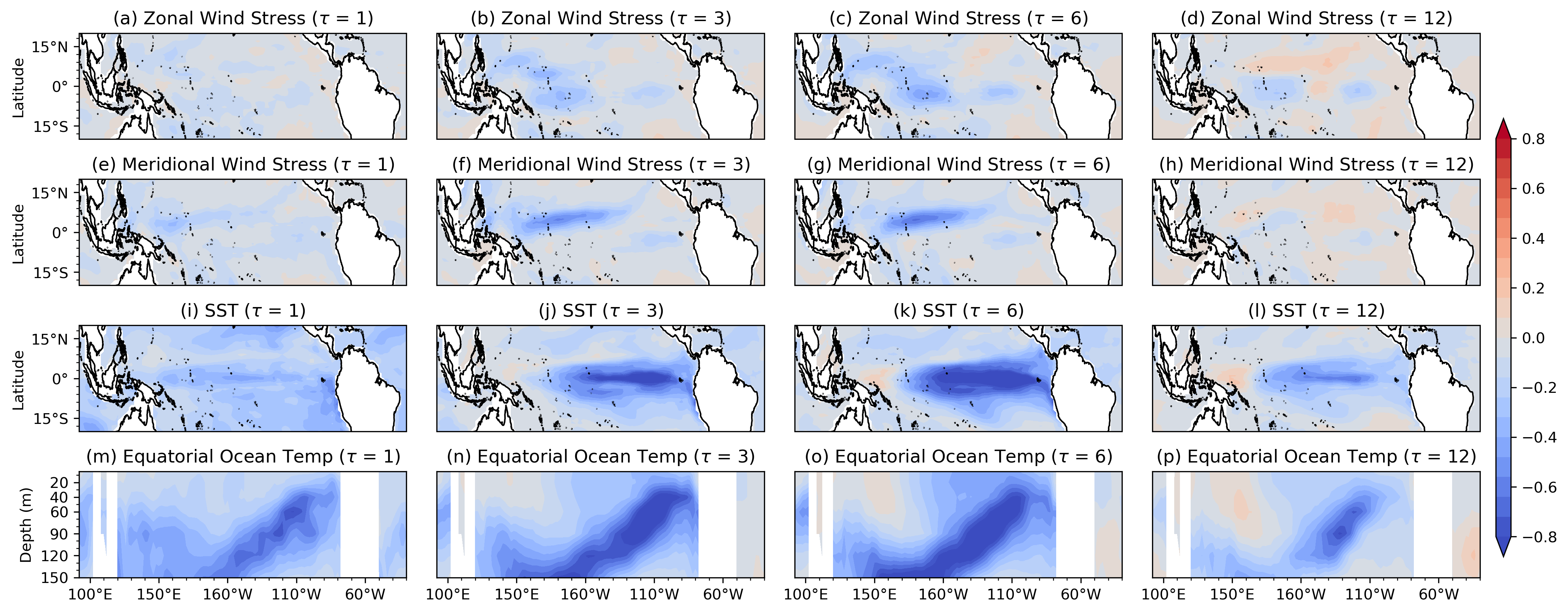}
    \caption{\textbf{Comparison of spatial patterns in normalized RMSE (normalization by the domain-averaged standard deviation of the corresponding variables) skill between Deep Learning and Linear Inverse Models as a function of lead time ($\tau$).} The first row (a–d) shows zonal-wind stress, the second row (e–h) meridional wind stress, the third row (i–l) Sea Surface Temperature (SST), and  fourth row (m–p) equatorial (5°N–5°S) ocean temperature.}
    \label{fig:predrmsediff}
\end{figure}

We now compare the forecast skill of DL, LIM and LIMOsst forecasts by initializing and verifying with the GODAS dataset. In terms of the domain-averaged correlation and RMSE metric of all variables, and the Nino3.4 Index, the DL forecasts consistently outperform both the LIM and LIMOsst forecasts across all variables and at all lead times (Figure \ref{fig:predictionskill}). Spatial maps of DL-forecast skill improvement reveal SST and ocean temperature spatial patterns similar to El-Niño (La-Niña) (Figs. \ref{fig:predcorrdiff} and \ref{fig:predrmsediff}). Specifically, from 1-month to 12-month lead time, the region with improved predictions evolves from off the equator to the equatorial region, accompanied by ocean temperature anomalies that tilt from the lower western to the upper eastern surface along the sloping thermocline. For surface wind stress, the DL model forecast-skill improvement is located in the central and western equatorial Pacific region, aligning with the region of improved skill for ocean temperature. This indicates that the DL model is able to better simulate the dynamics of ENSO compared to the LIMs.

Compared to the correlation metric, the improvement in RMSE by the DL model is not as significant, as illustrated in Figure \ref{fig:predictionskill}. An inherent advantage of the LIM at long lead times derives from the negative eigenvalues of $\mathbf{L}$, trending the forecasts towards zero anomaly as lead time increases. This means that the RMSE converges on the climatological standard deviation. In contrast, the DL model lacks this constraint and exhibits systematic errors that increase the RMSE, potentially to values larger than climatology.

It is noteworthy that in the equatorial western Pacific (near 135°E), DL forecast skill for SST and ocean temperature fields decreases with lead time. This appears to be a consequence of a systematic bias due to the exaggerated westward extension of the equatorial cold tongue in the CMIP6 models \citep{beobide-arsuagaUncertaintyENSOamplitudeProjections2021,jiangOriginsExcessiveWestward2021}. The Linear Inverse Model (LIM), trained exclusively on GFDL-CM4 data, demonstrates a mitigated version of this bias, owing to the relatively minor extent of the issue in GFDL-CM4 simulations. Conversely, the deep learning (DL) model, informed by a wider array of CMIP6 historical outputs, tends to accentuate the cold-tongue bias.

\section{Data Assimilation Results} \label{dl_res}

\begin{figure}
    \centering
    \includegraphics[width=1\textwidth]{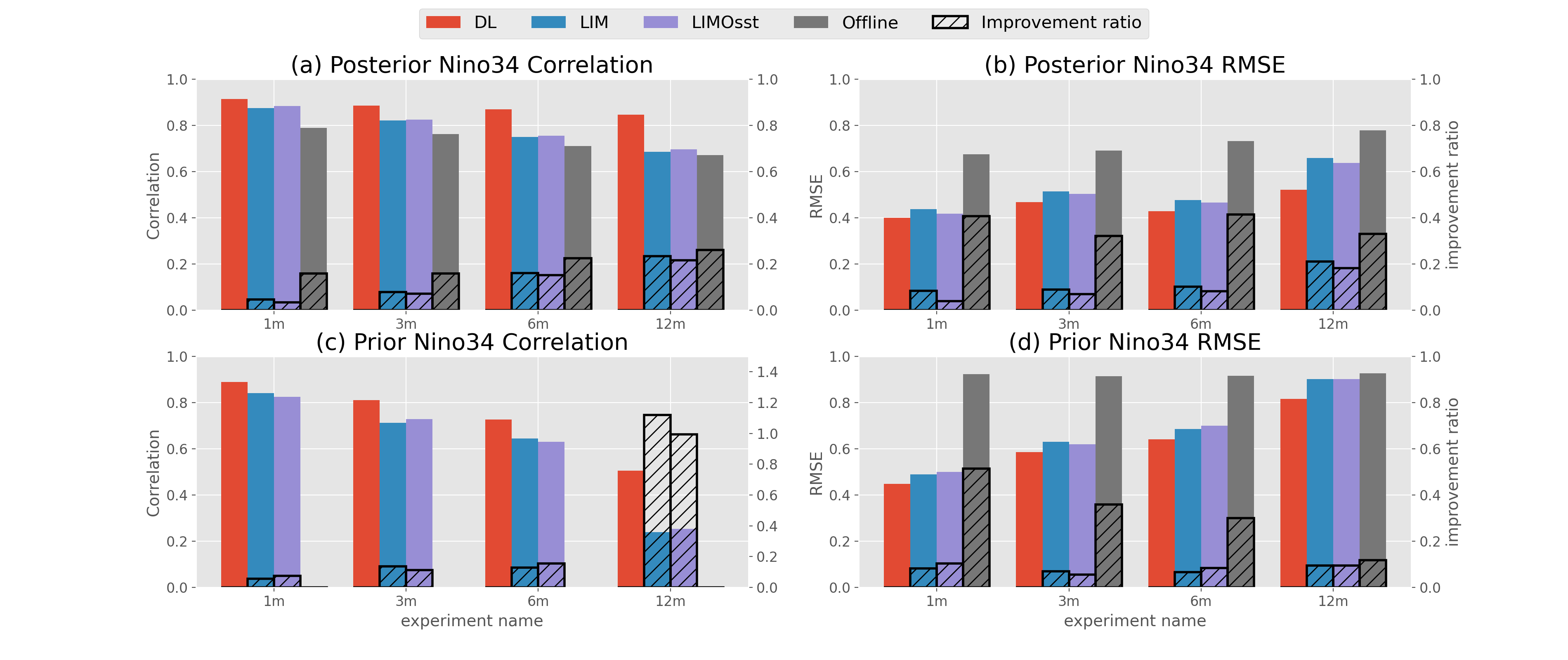}
    \caption{\textbf{Correlation (left panel, left $y$-axis) and RMSE (right panel, left $y$-axis) for the prior (lower panel)\& posterior (lower pannel) Nino3.4 index and GODAS dataset Nino3.4 index in 1-month, 3-month, 6-month and 12-month experiment.} Red bars show DL model result, blue bars are LIM result, the purple bars the LIMOsst model result, and gray bar the offline method result. The bars with diagonal hatching (right $y$-axis) shows the improvement ratio of the DL model compared to the LIM, LIMOsst model and offline method.}
    \label{fig:nino34_recon}
\end{figure}

\begin{figure}
    \centering
    \includegraphics[width=0.9\textwidth]{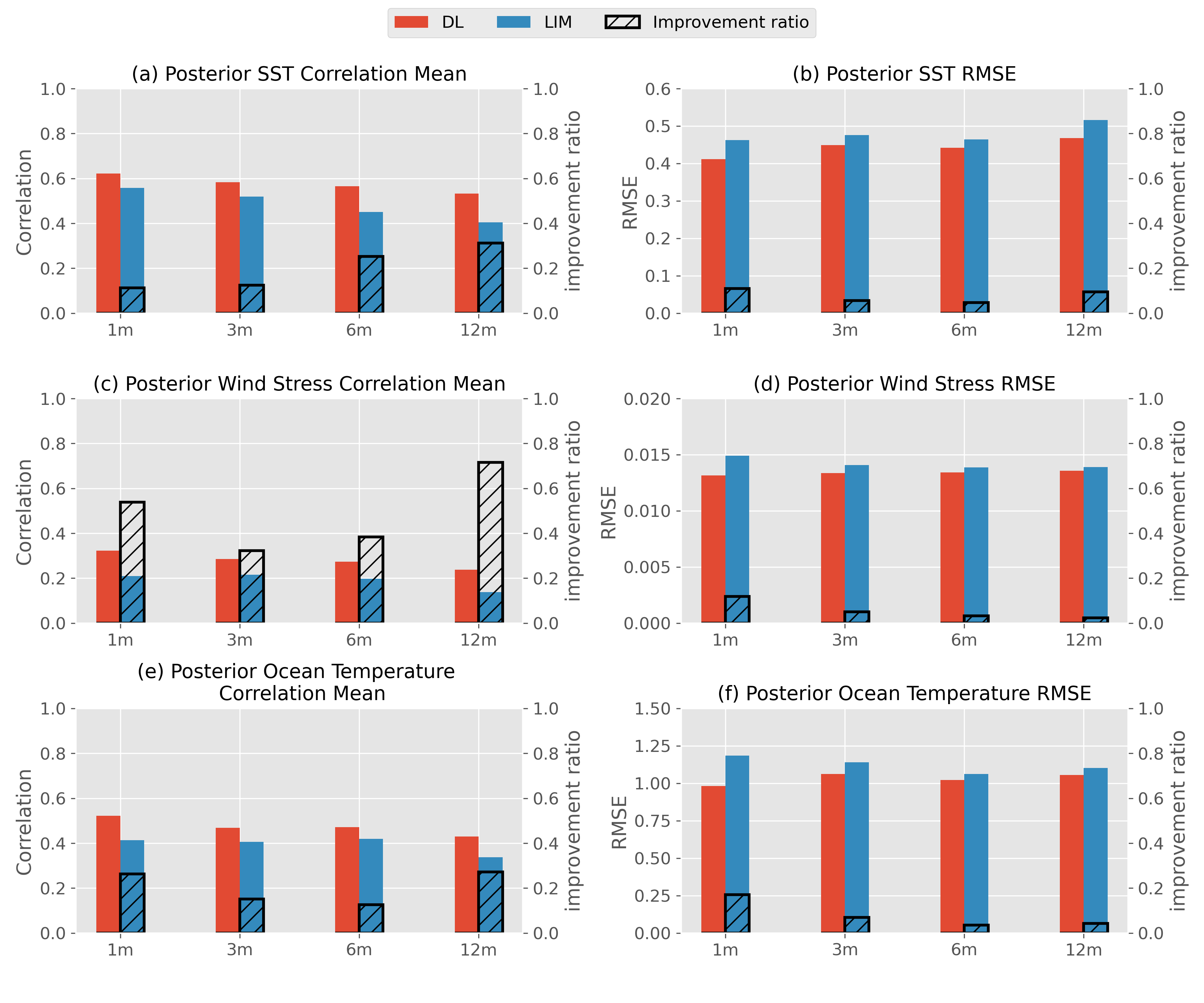}
    \caption{\textbf{Domain-averaged reconstructed (posterior) correlation and RMSE results for the Deep-learning model \& Linear Inverse Model verified on the GODAS dataset.} Red bars represent the Deep Learning (DL) model, blue bars  the Linear Inverse Model (LIM); bars with diagonal hatching indicate the improvement ratio of the results for the DL model relative to those for the LIM. The upper panel shows the SST field, the middle panel shows the wind stress field, and the lower panel shows the ocean temperature field.}
    \label{fig:posterior_compare}
\end{figure}

Assimilation experiments for a sparse network of noisy GODAS-sampled SST observations show that cycling with the DL model outperforms the others (LIM, LIMOsst, and offline) by around from 10 \% to 30 \% in reconstructing the Nino3.4 index (Fig. \ref{fig:nino34_recon}, top panels). Similar results are found for  skill in the prior forecast before data assimilation (Fig. \ref{fig:nino34_recon}, bottom panels). Improvement using the DL model increases with observation averaging time, which we attribute to the increase of forecast skill with lead time shown in section \ref{prediction_skill}. We find that these results are not sensitive to the SNR (see Supplementary Figure 2). In terms of skill across the entire domain, the DL model outperforms the LIMs for all observation averaging times and variables, most notably for correlation, and less so for RMSE Figure \ref{fig:posterior_compare}. Two key factors contribute to the modest improvement in the RMSE metrics. First, the enhancement in forecast skill in terms of RMSE is limited, as demonstrated in Fig. \ref{fig:predictionskill}. The RMSE improvement for SST, upper ocean temperature, and wind stress is relatively smaller compared to the improvement in correlation, as discussed in Section \ref{prediction_skill}. Secondly, as discussed previously, the LIM system is constrained by the fluctuation--dissipation relationship (\ref{eq:dcdt0}), which limits RMSE growth; the DL model does not have such constraint. Another possible contribution is the noise we have introduced to manage the loss of forecast variance and the limited number of ensemble members, since the DL model results improve modestly for larger ensembles (see supplementary Figure 3).

\begin{figure}
    \centering
    \includegraphics[width=1\textwidth]{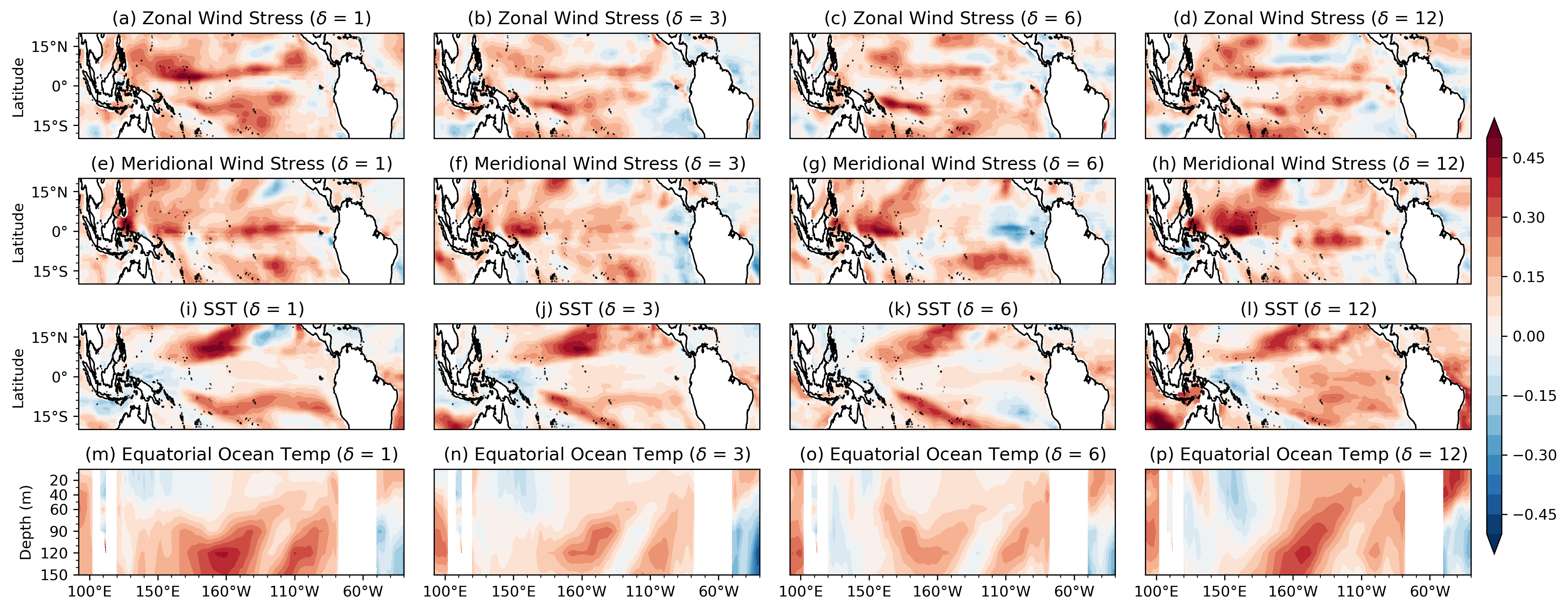}
    \caption{\textbf{The reconstructed result correlation skill difference between Deep-learning model and Linear Inverse Model of $\delta$–month experiment} in zonal wind stress (a–d), meridional wind stress (e–h), SST (i–l) and equatorial ocean temperature (from 5°N to 5°S) (m–p) field in 1,3,6 and 12 months-averaged experiments.}
    \label{fig:corr_diff}
\end{figure}

\begin{figure}
    \centering
    \includegraphics[width=1\textwidth]{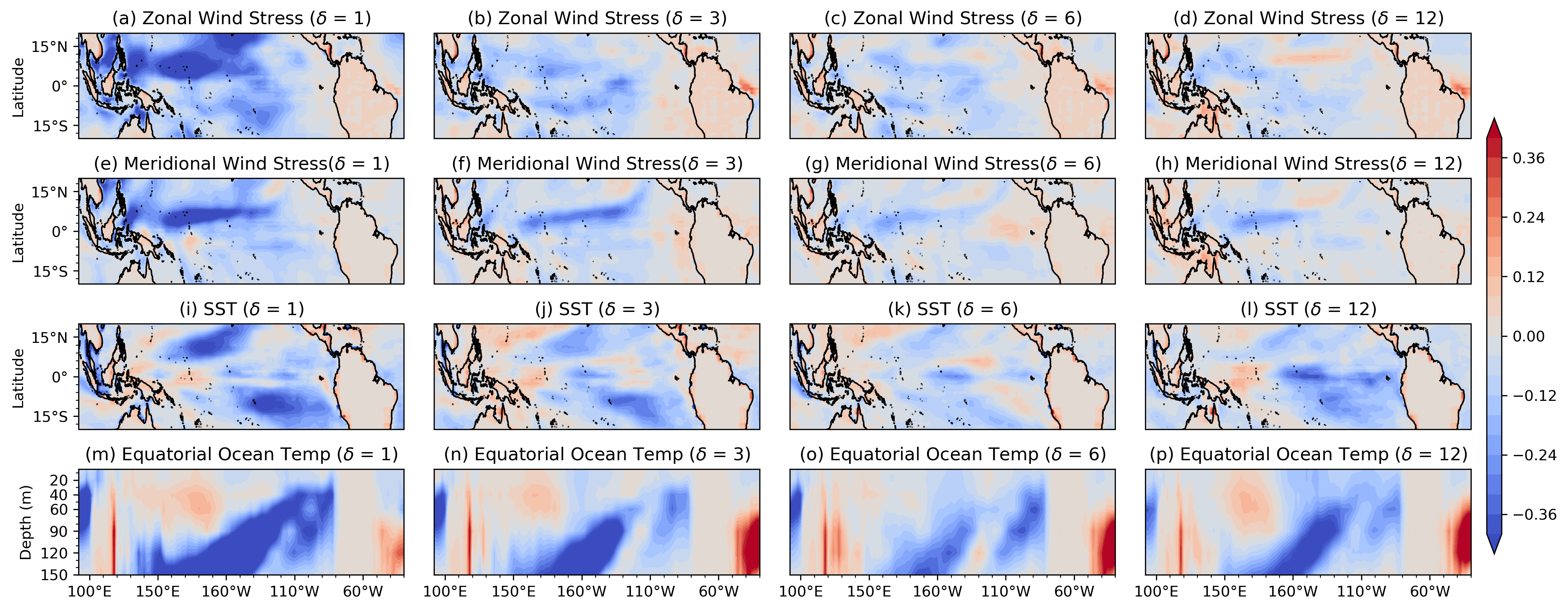}
    \caption{\textbf{The reconstructed, normalized RMSE (normalization by the domain-averaged standard deviation of the corresponding variables) skill difference between Deep-learning model and Linear Inverse Model of $\delta$–month experiment} in zonal wind stress (a–d), meridional wind stress (e–h), SST (i–l) and equatorial ocean temperature (from 5°N to 5°S) (m–p) field in 1,3,6 and 12-month experiment.}
    \label{fig:rmse_diff}
\end{figure}

The reconstruction correlation and RMSE spatial differences between DL and LIM is shown in the Figure \ref{fig:corr_diff} and \ref{fig:rmse_diff}, respectively. Improvement of the DL results over the LIM in zonal wind stress and SST are located primarily off the equator, which is not the same as for the forecasting experiments (cf. Figs. \ref{fig:predcorrdiff} and \ref{fig:predrmsediff}). This suggests that, relative to the LIM, the DL covariance estimates allow for more information extraction from the observations, which are located primarily closer to the equator. Improvements in the meridional wind stress are more closely confined to the equator relative to the forecasting experiments, suggesting a local influence of the observations. For the ocean temperature field, the reconstruction skill improvement of the DL model predominantly manifests in the mid-Pacific region at a depth of 100-150m, with extensions upward and eastward along the sloping thermocline. This spatial concentration roughly corresponds with the forecast skill improvements, especially for the 12-month observation averaging time. Areas where the DL model results are worse than the LIM are concentrated near South America and the land areas around the western Pacific warm pool, which corresponds with the forecast skill differences in RMSE and correlation (Figure \ref{fig:predrmsediff} and Figure \ref{fig:predcorrdiff}), but these areas are smaller in magnitude compared with skill enhancements elsewhere.

\begin{figure}
    \centering
    \includegraphics[width=1\textwidth]{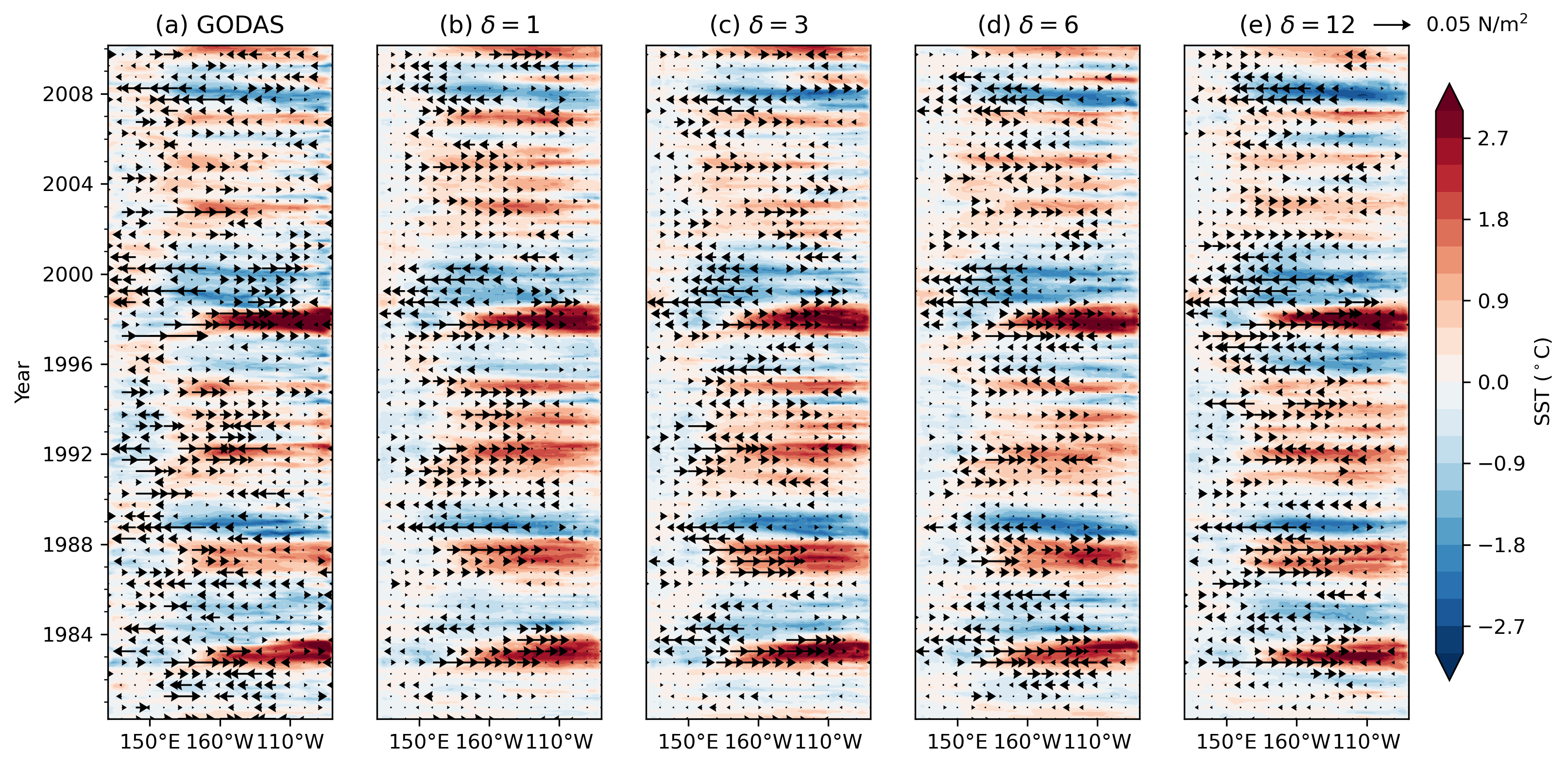}
    \caption{\textbf{The equatorial (5°N–5°S) evolution of SST (shading) and zonal wind stress anomalies (vectors) during 1980–2010 in $\delta$–month experiment.} (a) GODAS dataset. (b–e) DL-model reconstruction for experiments assimilating observations averaged over 1,3,6 and 12 months.}
    \label{fig:hm}
\end{figure}

Figure \ref{fig:hm} presents a comparison of the evolution of SST and zonal wind stress in the GODAS reanalysis and the DL data assimilation results. The reconstruction achieves remarkable accuracy, faithfully capturing the peaks and troughs of central-east Pacific SST as observed in the reanalysis, along with the corresponding zonal-wind stress. This is particularly notable in the 12-month averaged experiment, which utilizes only 24 observations per year, yet still largely captures the observed patterns. Notable differences include excessive easterly wind stress, and cooler SSTs, around 1996 and 2000, in the DL results when compared to GODAS.

\section{Conclusion and Discussion} \label{condis}

We have evaluated the potential of using a deep-learning model for cycling data assimilation on sparse observations to reconstruct the upper ocean and surface wind stress of the tropical Pacific ocean. The deep learning model is trained on CMIP6 model data following \citet{zhouSelfattentionBasedNeural2023}, and validated by forecasting on SODA reanalysis data. A significant drawback of the DL model for data assimilation is a bias for small forecast error variance. Therefore, we employed a approach to restore ensemble forecast variance by adding scaled samples from a library of DL model forecast errors to the DL model forecasts. We compare the performance of the DL model in forecasting and data assimilation experiments to control experiments using linear inverse models (LIMs) trained on the same CMIP6 dataset, and an offline data assimilation experiment that samples only from CMIP6 without a forecast model.

Overall, the results show that the DL model provides better forecasts compared to the LIM, especially in the central and eastern Pacific where ENSO dominates variability. The DL model outperformance is most notable in correlation (signal timing), and smaller in RMSE. For the data assimilation experiments, the results also show improvement using the DL model relative to the LIM, but the spatial distribution of these improvements differs from the forecasting results. In particular, relative to the LIM DA results, we find larger improvements off-equator in zonal-wind stress and SST, near the equator in ocean temperatures, and near the thermocline in the mid Pacific. These improvements reflect a combination of the forecast-skill improvements, which better retain the memory of past observations, and improved spatial covariance, which spread information from the sparse network of observations, which are more abundant over the equatorial western Pacific.

Based on this proof-of-concept study, we conclude that a deep-learning model can provide computationally efficient forecast priors for online paleoclimate data assimilation, leading to improved reconstruction outcomes. Future research will consider the application of these models to assimilating real proxy data, and extending the approach outside the tropics.

\pagebreak

\section*{Data and Code Availability} \label{sec:codedata}

GODAS dataset can be found: \url{https://psl.noaa.gov/data/gridded/data.godas.html}. SODA dataset can be found: \url{http://www.soda.umd.edu}. CMIP6 Dataset can be found: \url{https://pcmdi.llnl.gov/CMIP6/}. All codes, model weights and experimental results can be found: \url{https://github.com/ZiluM/DataAssimlationWithDL} after being accepted.

\section*{Acknowledgments}
GJH acknowledges support for this research from NSF awards 2202526 and 2105805, and Heising-Simons Foundation award 2023-4715. ZM acknowledges support for this research from NSF awards 2105805.

\bibliographystyle{ametsocV6}
\bibliography{main}  






\pagebreak

\section*{Supporting Information (SI)}

\setcounter{figure}{0}
\renewcommand{\thefigure}{S\arabic{figure}}

\begin{figure}
      \includegraphics[width=0.9\textwidth]{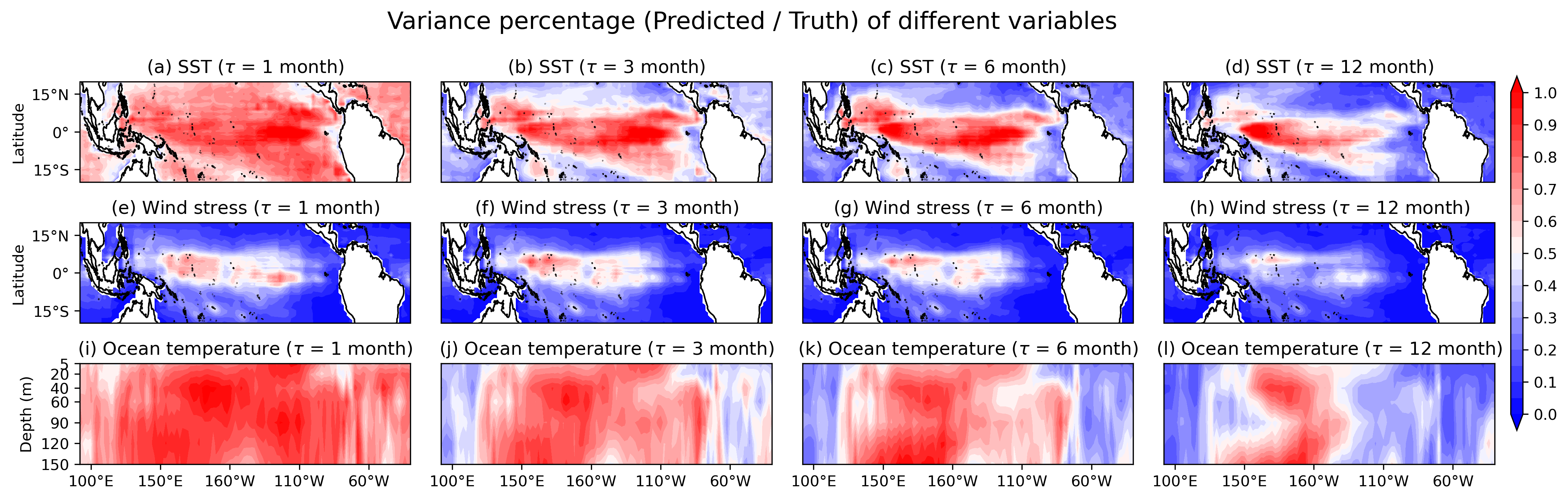}
      \caption{\textbf{Variance ratio of the Deep Learning forecasts to GODAS observations as a function of lead time, $\tau$,} for Sea Surface Temperature (SST) (upper panel), surface wind stress (middle panel), and ocean temperature (lower panel).}
      \label{sfig1}
\end{figure}

\begin{figure}
      \includegraphics[width=1\textwidth]{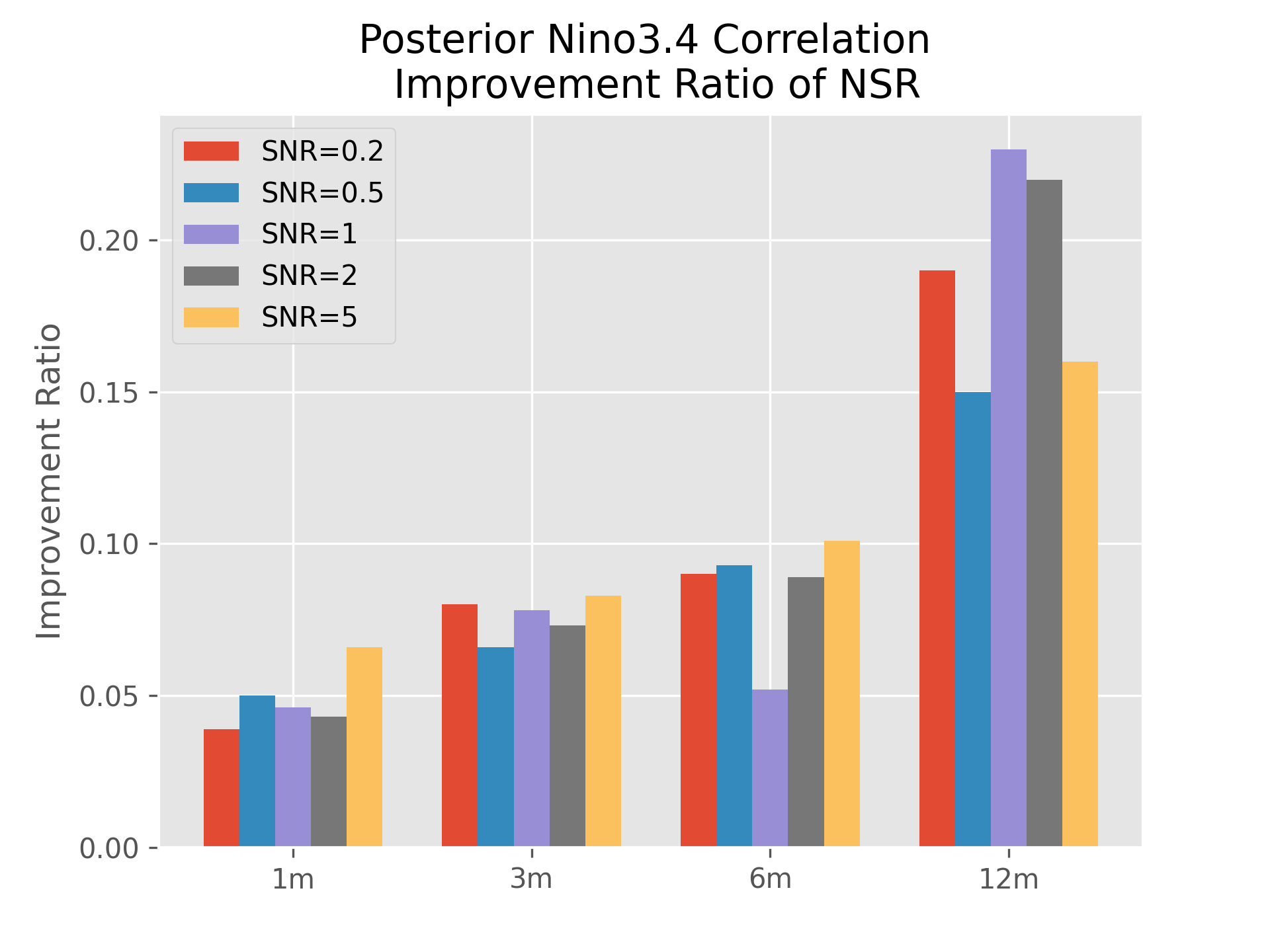}
      \caption{Improvement ratio of the Deep Learning model to LIM in data assimilation posterior results for the Nino3.4 index as a function of signal-to-noise ration (SNR).}
      \label{sfig2}
\end{figure}

\begin{figure}
      \includegraphics[width=1\textwidth]{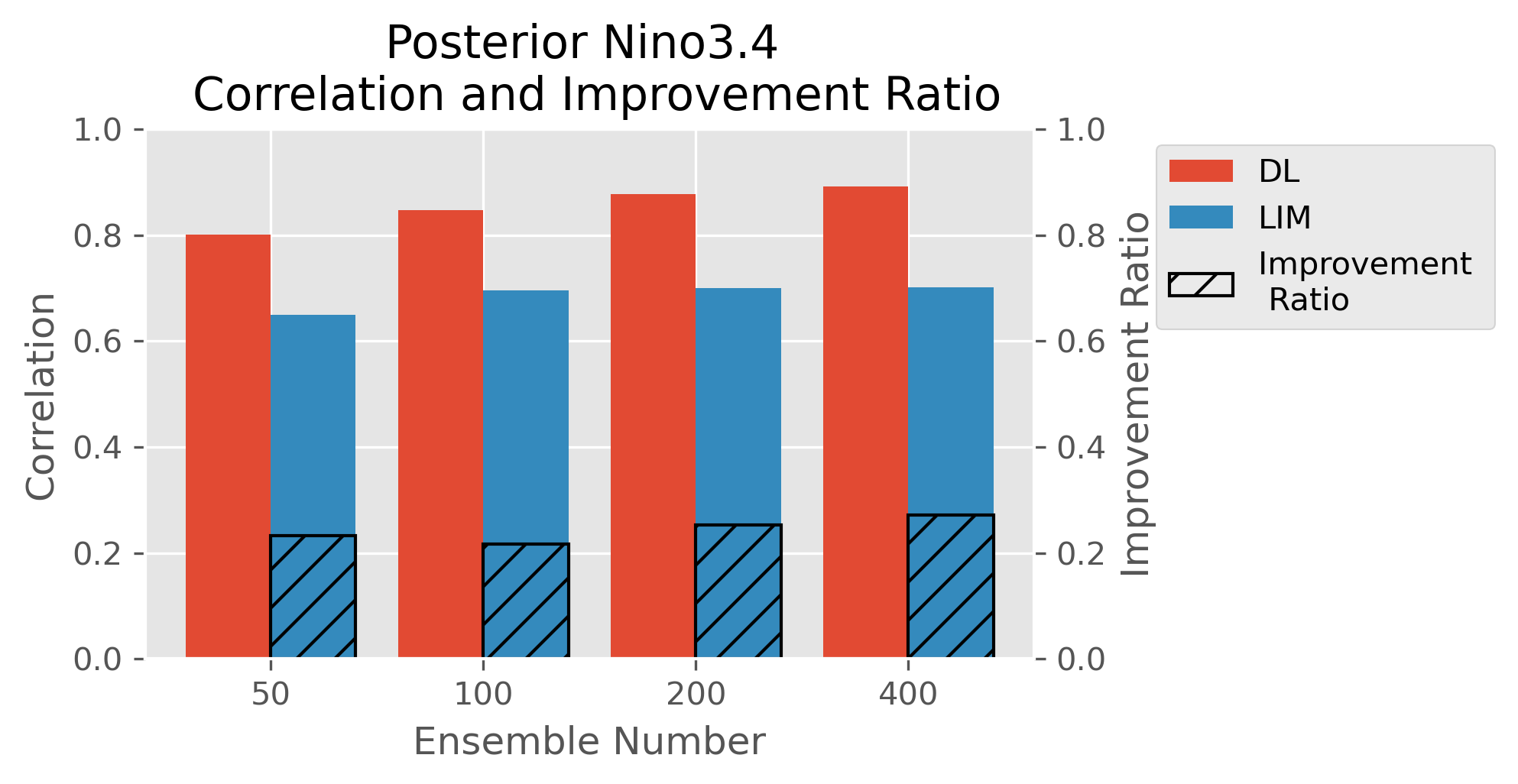}
      \caption{Correlation (left $y$-axis) and Improvement Ratio (right $y$-axis) for the Deep Learning model (red) and LIM (blue) reconstructed Niño 3.4 index in a 12-month-average experiment with SNR $=1$, plotted as a function of ensemble size.}
      \label{sfig3}
\end{figure}

As depicted in Figure \ref{sfig2}, the improvement ratio increases with observation averaging time, which likely derives from the increased predictive skill of the DL model. Nonetheless, there is no discernible trend across different Signal-to-Noise Ratios (SNR). A potential explanation for this is the ensemble size, which is fixed for all experiments at 100 members. DL-model performance improves slowly with increased ensemble size. For example, as illustrated in Figure \ref{sfig3}, while the correlation for the LIM remains unchanged in ensemble size, correlation for the DL-model results increases from 0.85 to 0.89 as the number of ensemble members rises from 100 to 400 for SNR=1 and 12-month-average experiments. This suggests that the accuracy of DL reconstructions can be modestly improved by expanding the ensemble size. Nevertheless, due to limitations in computational resources, all experiments are performed with 100 ensemble members, which yields results nearly as good as larger ensemble sizes.

\end{document}